\documentclass[prd,twocolumn,preprintnumbers,amsmath,amssymb,showpacs]{revtex4}

\usepackage{graphicx,bm}

\makeatletter
\def\graphicscale{\twocolumn@sw{0.3}{0.4}}
\def\graphicthreescale{\twocolumn@sw{0.3}{0.4}}

\begin{document}

\title{
The Landau-Ginzburg-Wilson approach to critical phenomena\\
in the presence of gauge symmetries
}

\author{Andrea Pelissetto}
\affiliation{Dipartimento di Fisica dell'Universit\`a di Roma ``La Sapienza"
        and INFN, Sezione di Roma I, I-00185 Roma, Italy}

\author{Antonio Tripodo, Ettore Vicari}
\affiliation{Dipartimento di Fisica dell'Universit\`a di Pisa
        and INFN, Sezione di Pisa, I-56127 Pisa, Italy}

\date{today}

\begin{abstract}
We critically reconsider the Landau-Ginzburg-Wilson (LGW) approach to
critical phenomena in the presence of gauge symmetries.  In the LGW
framework, to obtain the universal features of a continuous
transition, one identifies the order parameter $\Phi$ and considers
the corresponding most general $\Phi^4$ field theory that has the same
symmetries as the original model. In the presence of gauge symmetries,
one usually considers a gauge-invariant order parameter and a LGW
field theory that is invariant under the global symmetries of the
original model. We show that this approach, in which the gauge
dynamics is effectively integrated out, may sometimes lead to
erroneous conclusions on the nature of the critical behavior.  As an
explicit example, we show that the above-described LGW approach
generally fails for the three-dimensional ferromagnetic and
antiferromagnetic CP$^{N-1}$ models, which are invariant under global
U$(N)$ and local U(1) transformations.  We point out possible
implications for the finite-temperature chiral transition of nuclear
matter.

\end{abstract}



\maketitle


\section{Introduction}
\label{intro}

In the renormalization-group (RG) approach to critical phenomena, the
universal properties of continuous phase transitions can be obtained
by using the Landau-Ginzburg-Wilson (LGW) field-theoretical
approach~\cite{Landau-book,WK-74,Fisher-75,Ma-book,ZJ-book,PV-02}.  In
this framework the critical features are uniquely specified by the
nature of the order parameter associated with the critical modes, by
the symmetries of the model, and by the symmetries of the phases
coexisting at the transition, the so-called symmetry-breaking pattern.
In this paper we consider models which are also characterized by gauge
symmetries. In this case, the traditional LGW approach starts by
considering a gauge-invariant order parameter, effectively integrating
out the gauge degrees of freedom, and by constructing a LGW field
theory that is invariant under the global symmetries of the original
model.  We will show that in some cases this LGW approach leads to
erroneous conclusions on the nature of the critical behavior.

For this purpose, we consider the three-dimensional (3D) CP$^{N-1}$
model defined by the Hamiltonian
\begin{equation}
H_{\rm CP} = J \sum_{\langle {\bm x}{\bm y} \rangle} 
| \bar{\bm{z}}_{\bm x} \cdot {\bm z}_{\bm y} |^2,
\label{hcpn}
\end{equation}
where the sum is over the nearest-neighbor sites ${\langle {\bm x}{\bm
    y} \rangle}$ of a cubic lattice, and ${\bm z}_{\bm x}$ are
$N$-component complex vectors satisfying $\bar{\bm{z}}_{\bm x}\cdot
{\bm z}_{\bm x}=1$. The model is ferromagnetic for $J<0$ and
antiferromagnetic for $J > 0$.  CP$^{N-1}$ models have a global U($N$)
symmetry [${\bm z}_i \to U {\bm z}_i$ with a space-independent $U\in
  {\rm U}(N)$], and a local U(1) gauge symmetry (${\bm z}_{\bm x} \to
e^{i\theta_{\bm x}} {\bm z}_{\bm x}$).  The thermodynamics can be
derived from the standard partition function
\begin{equation}
Z=\sum_{ \{ {\bm z}_{\bm x}\}} \exp(-\beta H_{\rm CP}).
\label{zcpn}
\end{equation}
3D CP$^{N-1}$ models are expected to undergo a finite-temperature
transition between the high- and low-temperature phases. In the
ferromagnetic case, the order parameter may be identified as the
gauge-invariant site variable
\begin{equation}
Q_{{\bm x}}^{ab} = \bar{z}_{\bm x}^a z_{\bm x}^b - {1\over N} \delta^{ab},
\label{qdef}
\end{equation}
which is a hermitian and traceless $N\times N$ matrix. It
transforms as 
\begin{equation}
Q_{{\bm x}} \to U^\dagger Q_{{\bm x}} U,
\label{symmetry-U(N)}
\end{equation}
under global U($N$) transformations.  The order-parameter field in the
corresponding LGW theory is therefore a traceless hermitian matrix
field $\Phi^{ab}({\bm x})$, which can be formally defined as the
average of $Q_{\bm x}^{ab}$ over a large but finite lattice domain.
The LGW field theory is obtained by considering the most general
fourth-order polynomial in $\Phi$ consistent with the U($N$) symmetry
(\ref{symmetry-U(N)}):
\begin{eqnarray}
{\cal H} &=& {\rm Tr} (\partial_\mu \Phi)^2 
+ r \,{\rm Tr} \,\Phi^2 +  w \,{\rm tr} \,\Phi^3 \label{hlg}\\
&& +  \,u\, ({\rm Tr} \,\Phi^2)^2  + v\, {\rm Tr}\, \Phi^4 .
\nonumber
\end{eqnarray}
For $N=2$, the cubic term vanishes and the two quartic terms are
equivalent.  Therefore, one recovers the O(3)-symmetric LGW theory,
consistently with the equivalence between the CP$^1$ and the
Heisenberg model. For $N\ge 3$, the cubic term is generically expected
to be present. This is usually considered as the indication that phase
transitions of systems sharing the same global properties are of first
order, as one can easily infer using mean-field arguments. However, in
the large-$N$ limit, a different argument allows one to show that the
critical behavior of the ferromagnetic CP$^{N-1}$ model is the same as
that of an effective abelian Higgs model for an $N$-component complex
scalar field coupled to a dynamical U(1) gauge field~\cite{MZ-03}.
This equivalence is conjectured to extend to finite $N$ at the
critical point~\cite{MZ-03}.  The RG flow of the abelian Higgs model
presents a stable fixed point for a sufficiently large number of
components~\cite{HLM-74}.  Thus, for large values of $N$, 3D
CP$^{N-1}$ models may undergo a continuous transition, in the same
universality class as that occurring in the abelian Higgs model,
contradicting the LGW results.  The predictions derived from the LGW
Hamiltonian (\ref{hlg}) are also contradicted by recent numerical
studies \cite{NCSOS-11,NCSOS-13}, which provide evidence of continuous
transitions in models that are expected to be in the same universality
class as that of the 3D CP$^2$ model.  All these results suggest that
the critical modes at the transition are not exclusively associated
with the gauge-invariant order parameter $Q$ defined in
Eq.~(\ref{qdef}). Other features, for instance the gauge degrees of
freedom, become relevant, requiring an effective description different
from that of the LGW theory (\ref{hlg}).

In this paper, we show that the LGW approach, in which the gauge
degrees of freedom are somehow integrated out, also fails for the
antiferromagnetic CP$^{N-1}$ model (ACP$^{N-1}$), i.e., in model
(\ref{hcpn}) with $J>0$, for $N=4$. In the antiferromagnetic case the
same argument \cite{DPV-15} used for $J < 0$ allows one to identity
the order parameter with a staggered version of the site variable
$Q^{ab}_x$. In the LGW approach, the fundamental field is therefore a
hermitean traceless matrix $\Phi^{ab}(x)$ and the corresponding field
theory is given by Eq.~(\ref{hlg}). At variance with the ferromagnetic
case, in the ACP$^{N-1}$ case there is also a discrete ${\mathbb Z}_2$
symmetry that forces $w = 0$. For $N=2$ and $N=3$ this approach
predicts a transition in the same universality class as that of the
$O(n)$ vector model, with $n=3$ and 8, respectively.  The result for
$N=2$ is consistent with the exact mapping between CP$^1$ and vector
O(3) models.  The prediction for $N=3$ has been verified
numerically~\cite{DPV-15}: for this value of $N$ the standard LGW
approach provides the correct description of the critical modes.  For
$N\ge 4$, the analysis of the RG flow in the effective LGW theory does
not identify stable fixed points that can be associated with
continuous transitions~\cite{DPV-15}. Therefore, the effective theory
predicts that possibly present transitions are of first order.  The
numerical results we report here contradict this prediction.  For the
ACP$^3$ model, we have numerical evidence of a continuous
transition. Again, the effective theory for a gauge-invariant order
parameter does not provide a correct description of the critical
behavior.

Our considerations may be relevant for the finite-temperature
transition of nuclear matter between the low-temperature hadronic
phase, in which chiral symmetry is broken, and the high-temperature
quark-gluon phase, in which chiral symmetry is restored in the limit
of massless
quarks~\cite{PW-84,Wilczek-92,RW-93,GGP-94,Wilczek-00,Karsch-02}.  The
nature of this transition is still controversial, in particular in the
case of two light flavors, in spite of several Monte Carlo studies
using different lattice QCD
formulations~\cite{CP-PACS-01,MILC-00,KLP-01,KS-01,EHMS-01,DDP-05,KS-06,
  FP-07,Bazetal-12,Burger-etal-13,BDDPS-14,DD-17}.  Some studies favor
a continuous transition, but are not sufficiently accurate to clearly
identify the corresponding universality class, while others report
evidence of a first-order transition.

Our understanding of the QCD finite-temperature phase transition in
the presence of $N_f$ light quarks is based on the analysis of the LGW
effective theory.  The relevant order-parameter field \cite{PW-84} is
an $N_f\times N_f$ complex-matrix field $\Phi_{ij}$, related to the
bilinear quark operators $\bar{\psi}_{Li}\psi_{Rj}$, which is the
analogue of the bilinear $Q^{ab}$ of the CP$^{N-1}$ models.  The
corresponding LGW theories have been analyzed in detail and the
corresponding predictions for the nature of the transitions have been
extensively discussed~\cite{PW-84,BPV-03,BPV-05,v-07,PV-13}.  Note
that they are all based on the assumption that the relevant critical
modes are only associated with the local, gauge-invariant bilinear
fermion operators. In particular, it is implicitly assumed that gauge
modes play no role at the transition, a hypothesis that should not be
taken for granted, as the examples discussed in this paper indicate.

The paper is organized as follows.  In Sec.~\ref{arpcp} we define the
LGW $\Phi^4$ theory which is supposed to describe the universal
properties of the critical transition in the ACP$^{N-1}$ model and
briefly review its predictions \cite{DPV-15}.  In Sec.~\ref{acp3res}
we discuss the results of Monte Carlo simulations of the ACP$^3$
lattice model, which provide evidence of a continuous
transition. Finally, in Sec.~\ref{conclu} we summarize our
conclusions.

\section{The LGW $\Phi^4$ theory for the ACP$^{N-1}$ models}
\label{arpcp}

In this section we review the derivation of the LGW theory for the
ACP$^{N-1}$ model \cite{DPV-15}, emphasizing the main assumptions.
Similar arguments \cite{FMSTV-05} also hold for the antiferromagnetic
RP$^{N-1}$ model, in which spins are real.

As a first step we should identify the order parameter. In the
antiferromagnetic case we expect a breaking of translational
invariance in the low-temperature phase. Therefore, a global order
parameter is the staggered quantity
\begin{equation}
A^{ab} = 
       \sum_{\bm x} p_{\bm x} Q_{\bm x}^{ab} = 
       \sum_{\bm x} p_{\bm x} \bar{z}^a_{\bm x} z^b_{\bm x},
\label{biab}
\end{equation}
where $p_{\bm x}$ is the parity of the site ${\bm x} \equiv
(x_1,x_2,x_3)$ defined by $p_{\bm x} = (-1)^{\sum_k x_k}$. The matrix
$A^{ab}$ is hermitian and traceless.  Moreover, it changes sign under
translations of one site, which exchange the even- and odd-parity
sublattices.  In order to construct the LGW model, we replace $A$ with
a local variable $\Phi(x)$ which is taken as fundamental variable
(essentially, one imagines that $\Phi$ is defined as $A$, but now the
summation extends only over a large, but finite, sublattice).  Then,
the corresponding LGW theory is obtained by writing down the most
general fourth-order polynomial that is invariant under U($N$)
transformations and under the global ${\mathbb Z}_2$ transformation
$\Phi \to -\Phi$, i.e.
\begin{eqnarray}
{\cal H}_a = {\rm Tr} (\partial_\mu \Phi)^2 + r \,{\rm Tr} \,\Phi^2 +
{u_0\over 4} \, ({\rm Tr} \,\Phi^2)^2 + {v_0 \over 4} \, {\rm Tr}\,
\Phi^4 .
\label{ahlg} 
\end{eqnarray}
Note that the symmetry $\Phi\to-\Phi$ does not hold in ferromagnetic
CP$^{N-1}$ models, so that the corresponding LGW theory (\ref{hlg})
contains also a cubic term.

Since any $2\times 2$ and $3\times3$ traceless matrix $\Phi$
satisfies
\begin{equation}
{\rm Tr}\, \Phi^4 = {1\over 2} ({\rm Tr} \,\Phi^2)^2,
\label{eq2e3}
\end{equation}
the two quartic terms of the Hamiltonian (\ref{ahlg}) are equivalent
for $N=2$ and $N=3$. Therefore, the $N=2$ and $N=3$ $\Phi^4$ theories
(\ref{ahlg}) can be exactly mapped onto the O(3) and O(8) symmetric
$\Phi^4$ vector theories, respectively.  This implies that the
continuous transitions of the ACP$^1$ and ACP$^2$ models belong to the
O(3) and O(8) vector universality classes, respectively.  For $N=3$
this is a highly nontrivial prediction, as it implies a dynamical
enlargement of the symmetry at the critical point.  The O(8) nature of
the transition has been confirmed~\cite{DPV-15} by a detailed Monte
Carlo (MC) study of the ACP$^2$ model. Therefore, for this value of
$N$, the LGW theory provides the correct effective description of the
critical behavior.

The nature of the transitions in ACP$^{N-1}$ models for $N\ge 4$ has
been investigated by analyzing the RG flow of the LGW theory
(\ref{ahlg}) in the space of the two quartic parameters, using two
different perturbative methods \cite{DPV-15}. The analysis of the
five-loop series in the ${\overline {\rm MS}}$ renormalization
scheme~\cite{TV-72} and of the six-loop series in the massive
zero-momentum scheme~\cite{Parisi-80,ZJ-book,PV-02} both indicate the
absence of stable fixed points for $N=4$ and $N=6$. As a consequence,
the LGW approach predicts the absence of continuous transitions for
these two values of $N$.  However, as we shall see, this is
contradicted by the numerical results for $N=4$ presented in the next
section, which provide a robust evidence of a continuous transition.

\section{Numerical results for the ACP$^3$ lattice model}
\label{acp3res}

In this section we present a numerical investigation of the ACP$^3$
lattice model (\ref{hcpn}).  We set $J=1$ and perform MC simulations
of cubic systems of linear size $L$ with periodic boundary
conditions. Because of the antiferromagnetic nature of the model we
only consider even values of $L$, up to $L=48$.  We use a standard
Metropolis algorithm \cite{footnoteMC}.

In our MC simulations we compute correlations of the gauge invariant
operator $Q_{\bm x}^{ab}$ defined in Eq.~(\ref{qdef}).  Its two-point
correlation function is defined as
\begin{equation}
G({\bm x}-{\bm y}) = \langle {\rm Tr}\, Q_{\bm x}^\dagger  
Q_{\bm y} \rangle  
\label{gxyp}
\end{equation}
Due to the staggered nature of the ordered parameter, we only consider
correlations between points that have the same parity.  The
susceptibility and the correlation length are defined as
\begin{eqnarray}
&&\chi =  \sum_{{\rm even} \; {\bm x}} G({\bm x}) = 
\widetilde{G}({\bm 0}), 
\label{chisusc}\\
&&\xi^2 \equiv  {1\over 4 \sin^2 (p_{\rm min}/2)} 
{\widetilde{G}({\bm 0}) - \widetilde{G}({\bm p})\over 
\widetilde{G}({\bm p})},
\label{xidefpb}
\end{eqnarray}
where $\widetilde{G}({\bm p})=\sum_{{\rm even} \; {\bm x}} e^{i{\bm
    p}\cdot {\bm x}} G({\bm x})$ is the Fourier transform of $G({\bm
  x})$ over the even-parity sublattice, ${\bm p} = (p_{\rm min},0,0)$,
and $p_{\rm min} \equiv 2 \pi/L$.  Finally, we consider the Binder
parameter
\begin{equation}
U = { \langle [\sum_{{\rm even} \; {\bm x}} {\rm Tr}\, Q_{{\bm 0}}^\dagger  
Q_{\bm x} ]^2 \rangle \over  
\langle \sum_{{\rm even} \; {\bm x}} {\rm Tr}\, Q_{{\bm 0}}^\dagger  
Q_{\bm x} \rangle^2 } .
\label{binderdef}
\end{equation}
\begin{figure}[tbp]
\includegraphics*[scale=\graphicscale]{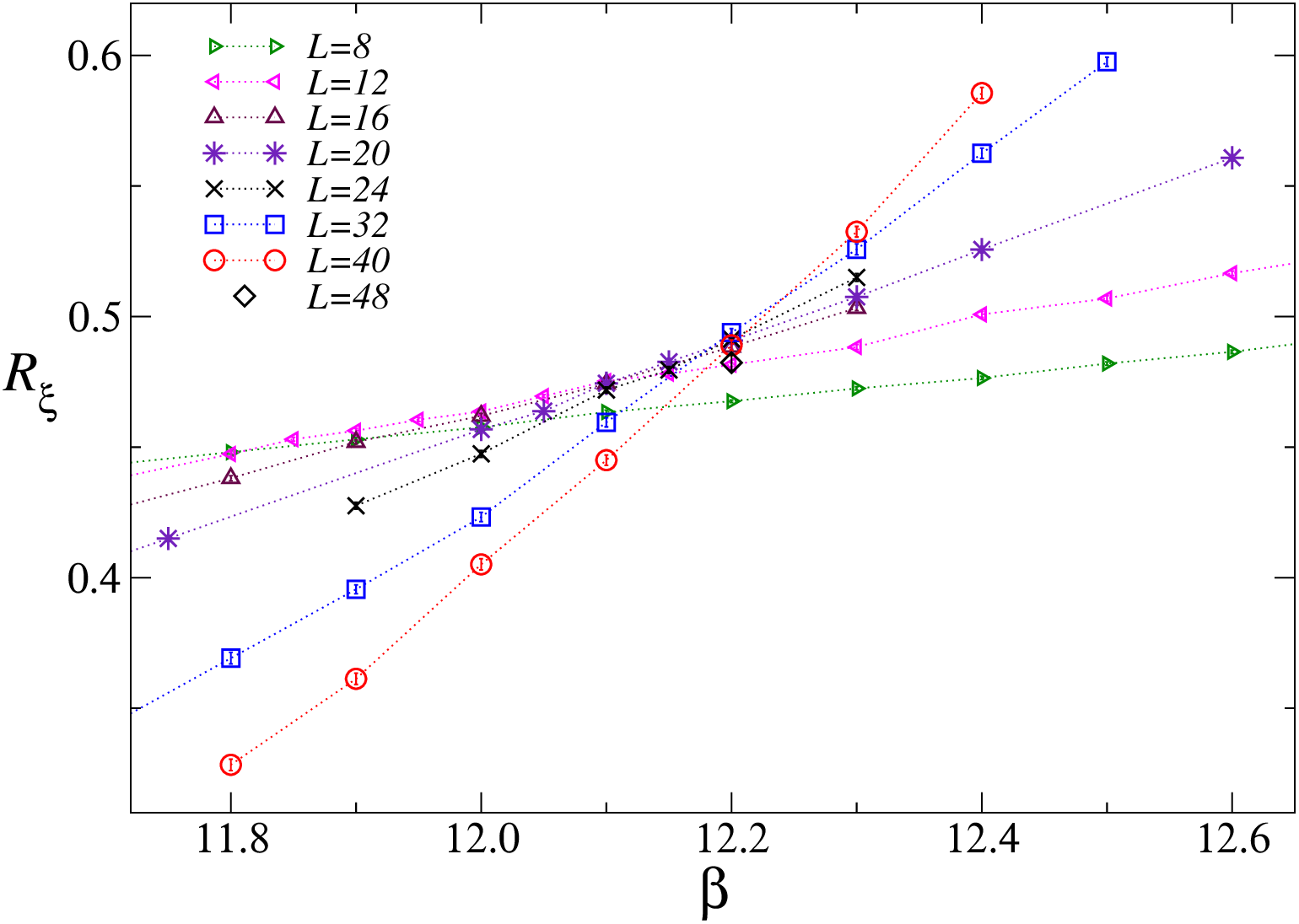}
\caption{MC data of $R_\xi$ for the ACP$^3$ lattice model and several
  lattice sizes $L$ up to $L=48$.  The data sets corresponding to
  different values of $L$ cross for $\beta \approx 12.2$.  The dotted
  lines are only meant to guide the eye.}
\label{rxin4}
\end{figure}
To determine the critical behavior we study the finite-size behavior.
The finite-size scaling (FSS) limit is obtained by taking $\beta\to
\beta_c$ and $L\to\infty$ keeping
\begin{equation}
X \equiv (\beta-\beta_c)L^{1/\nu}
\label{Xdef}
\end{equation}
fixed, where $\beta_c$ is the inverse critical temperature and $\nu$
is the correlation-length exponent. Any RG invariant quantity $R$,
such as $R_\xi\equiv \xi/L$ and $U$, is expected to asymptotically
behave as
\begin{eqnarray}
R(\beta,L) = f_R( X ) + L^{-\omega} g_R(X) + \ldots
\label{rsca}
\end{eqnarray}
where $f_R(X)$ is a universal function apart from a trivial
normalization of the argument. In particular, the quantity $R^* \equiv
f_R(0)$ is universal within the given universality class.  The
approach to the asymptotic behavior is controlled by the universal
exponent $\omega>0$, which is associated with the leading irrelevant
RG operator.  

To identify a transition point, we check whether the estimates of
$R_\xi \equiv \xi/L$ and of $U$ for different values of $L$ have a
crossing point as a function of $\beta$. The MC estimates of $R_\xi$
reported in Fig.~\ref{rxin4} clearly show a crossing point, providing
evidence of a transition at $\beta \approx 12.2$.  The Binder
parameter $U$ behaves analogously.

\begin{figure}[tbp]
\includegraphics*[scale=\graphicscale]{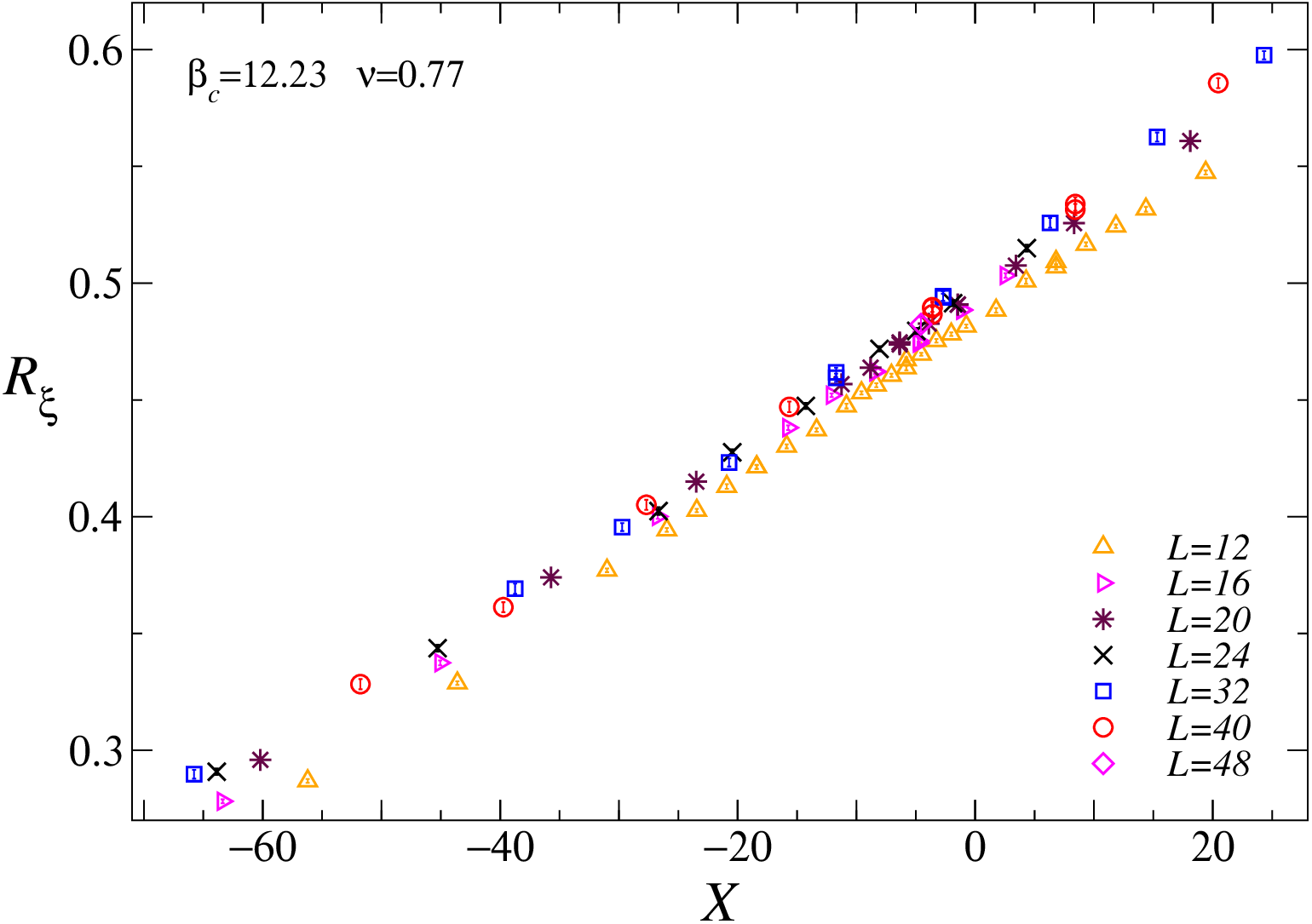}
\caption{ $R_\xi$ versus $X\equiv L^{1/\nu}(\beta-\beta_c)$ with
  $\beta_c=12.23$ and $\nu=0.77$. The data approach a scaling curve
  with increasing $L$, supporting the scaling behavior (\ref{rsca}).
}
\label{rxin4sca}
\end{figure}

\begin{table}
\caption{Results of the fits, for different choices of 
$\Delta$, $L_{\rm min}$, and $n$. For a given $\Delta$ and $L_{\rm min}$ only
results corresponding to $\beta$ and $L$ such that 
$|R_{\xi}(\beta,L) - R_{\xi}^*| < \Delta$ and $L\ge L_{\rm min}$ 
are included in the fit (we set $R_\xi^* = 0.50$).
The parameter $n$ is the order of the polynomial used in the fit,
Eq.~(\ref{fitr}). In the combined fits the order
is the same for both observables. For each fit we report the sum of the 
residuals ($\chi^2$) divided by the number of degrees of freedom (DOF) of the 
fit, $\beta_c$, and $\nu$.
}
\label{tab:fit}
\begin{tabular}{ccccll} 
\hline\hline
$\Delta$ & $L_{\rm min}$ & $n$ & $\chi^2$/DOF & 
\multicolumn{1}{c}{$\beta_c$} & 
\multicolumn{1}{c}{$\nu$} \\
\hline
\multicolumn{6}{c}{Combined analysis}   \\ 
0.20  &  24 &   1 &    7.5 & 12.224(4) &   0.78(1)  \\
0.20  &  24 &   2 &    2.4 & 12.229(4) &   0.78(2)  \\
0.20  &  24 &   3 &    2.3 & 12.229(5) &   0.79(2)  \\
0.20  &  32 &   1 &    5.2 & 12.249(8) &   0.87(4)  \\
0.20  &  32 &   2 &    1.7 & 12.246(8) &   0.79(4)  \\
0.20  &  32 &   3 &    1.7 & 12.247(8) &   0.78(4)  \\
0.10  &  24 &   1 &    3.9 & 12.225(4) &   0.78(2)  \\
0.10  &  24 &   2 &    2.6 & 12.229(5) &   0.79(3)  \\
0.10  &  32 &   1 &    2.2 & 12.247(8) &   0.85(6)  \\
0.10  &  32 &   2 &    1.8 & 12.246(8) &   0.82(6)  \\
0.05  &  24 &   1 &    3.4 & 12.230(6) &   0.80(7)  \\
0.05  &  24 &   2 &    3.0 & 12.229(6) &   0.75(7)  \\
0.05  &  32 &   1 &    1.6 & 12.262(16)&   0.81(16)  \\
0.05  &  32 &   2 &    1.5 & 12.248(16)&   0.60(14)  \\
\multicolumn{6}{c}{Analysis of $U$}   \\
0.10  &  24 &   1 &    2.2 & 12.34(2)  &   0.99(11) \\
0.10  &  32 &   1 &    1.7 & 12.31(2)  &   0.62(8)  \\
0.05  &  24 &   1 &    1.1 & 12.31(2)  &   0.68(13) \\
0.05  &  32 &   1 &    1.4 & 12.38(10) &   0.89(48) \\
\multicolumn{6}{c}{Analysis of $R_\xi$}   \\
0.10  &  24 &   1 &    2.4 & 12.211(4) &   0.74(2) \\
0.10  &  32 &   1 &    1.7 & 12.24(1)  &   0.86(6) \\
0.05  &  24 &   1 &    1.6 & 12.216(6) &   0.77(7) \\
0.05  &  32 &   1 &    1.4 & 12.25(1)  &   0.77(16) \\
\hline\hline
\end{tabular}
\end{table}

To obtain a quantitative estimate of $\beta_c$ and of the exponent
$\nu$, we perform nonlinear fits of $R_\xi$ and $U$ around the
crossing point. We first use the simple Ansatz
\begin{equation}
R = R^* + c_1\, L^{1/\nu}\,(\beta - \beta_c),
\label{rxisa}
\end{equation}
for $R = R_\xi$ and $U$. This Ansatz apparently describes well the
data in a relatively large interval around the transition point,
essentially when $\Delta = |R_\xi-R_\xi^*| \lesssim 0.10 $.  We have
also performed fits considering a second-order and a third-order
polynomial in $X$, i.e., fitting $R$ to 
\begin{equation}
R = R^* + \sum_{k=1}^n c_k X^k,
\label{fitr}
\end{equation}
with $n=2$ and $n=3$.
Finally, we also performed combined fits of $R_\xi$
and $U$. The data are not sufficiently precise to allow us to include
scaling corrections in the fit. Therefore, to estimate their
relevance, we have repeated all fits several times, each time only
including data satisfying $L \ge L_{\rm min}$, varying $L_{\rm min}$
between 8 and 32. Some results are reported in Table~\ref{tab:fit}.
The analyses of the Binder parameter give estimates with somewhat large errors.
Moreover, the estimates of $\nu$ show a significant scatter as $L_{\rm min}$
and $\Delta$ are varied. Fits of $R_\xi$ are more stable. To obtain the 
final estimates, we consider the combined fits, that give reasonably accurate 
and stable results.  We finally estimate
\begin{eqnarray}
\beta_c=12.23(6),\qquad \nu = 0.77(5),\label{fitres}
\end{eqnarray}
and $R_\xi^* = 0.50(1)$, $U^* = 1.025(1)$, where the errors take into
account the dependence of the results on the range $\Delta$, on
$L_{\rm min}$ and on the order $n$ of the polynomial.  In
Fig.~\ref{rxin4sca} we show a plot of $R_\xi$ versus $X$, using the MC
estimates of $\beta_c$ and $\nu$. As $L$ increases, data collapse onto
a single scaling curve.

\begin{figure}[tbp]
\includegraphics*[scale=\graphicscale]{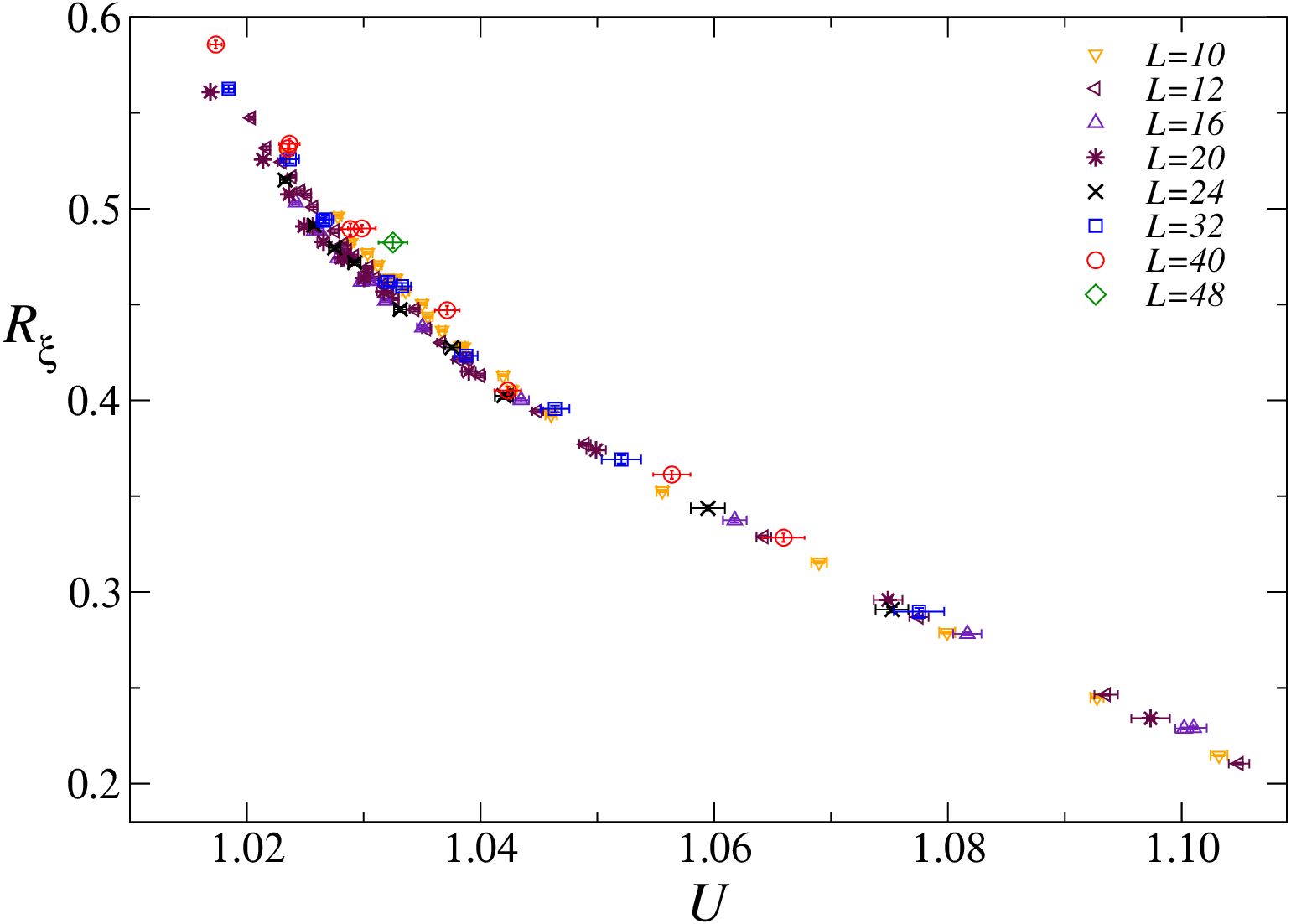}
\caption{Plot of $R_\xi$ vs $U$.  The data approach a scaling curve
  with increasing $L$, supporting the scaling behavior
  (\ref{rxibisca}).  }
\label{rxibin4}
\end{figure}

Since both $R_\xi$ and $U$ satisfy Eq.~(\ref{rsca}), we must have
\begin{equation}
R_\xi =  F(U) + O(L^{-\omega}),
\label{rxibisca}
\end{equation}
where $F(U)$ is a universal function.  In Fig.~\ref{rxibin4} we plot
$R_\xi$ versus $U$. The data collapse onto a single curve without the
need of tuning any parameter, confirming the scaling behavior
(\ref{rxibisca}). Scaling corrections are smaller for $\beta<\beta_c$
than for $\beta > \beta_c$.

\begin{figure}[tbp]
\includegraphics*[scale=\graphicscale]{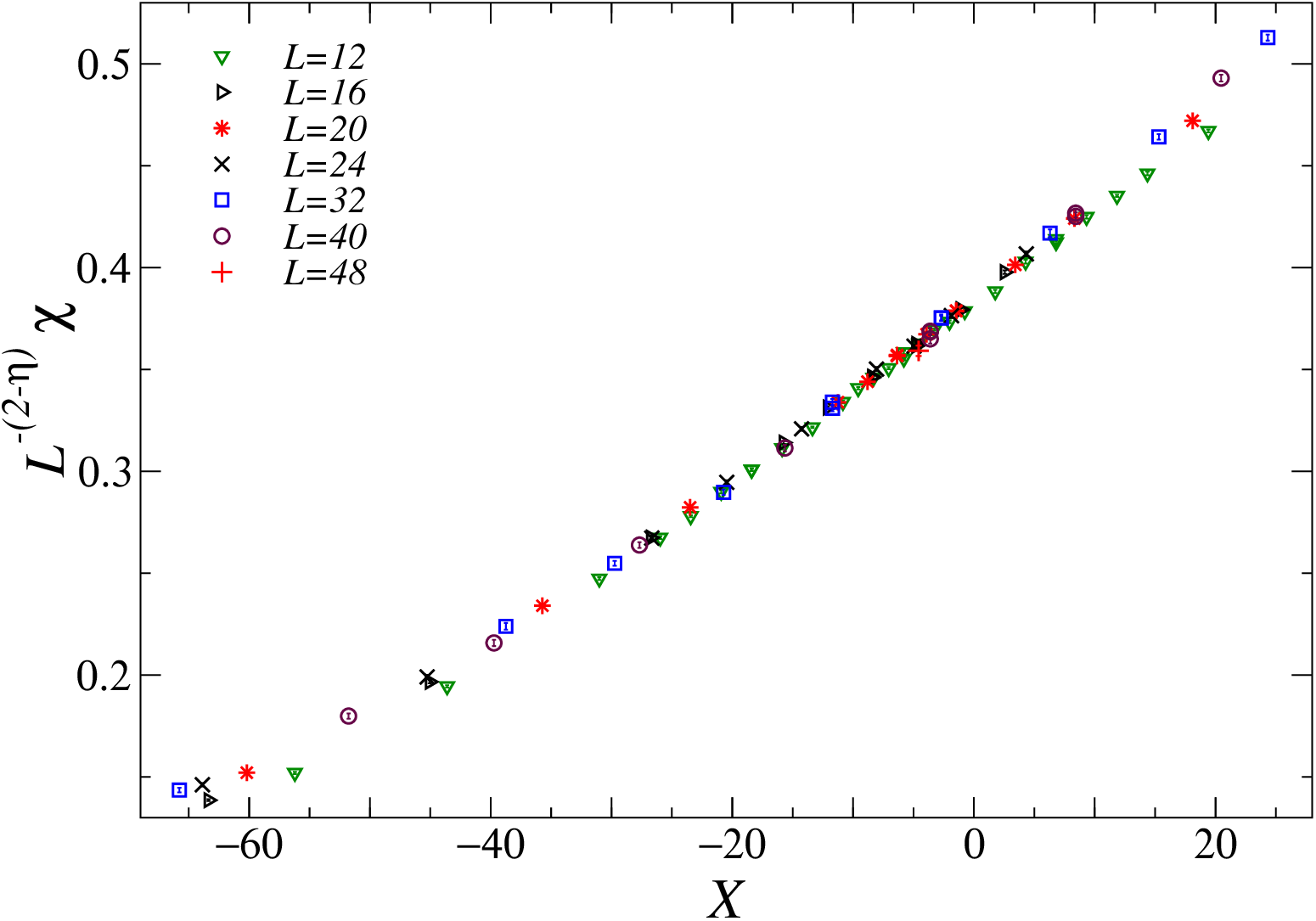}
\caption{ We plot $\chi/L^{2-\eta}$ versus $X\equiv
  L^{1/\nu}(\beta-\beta_c)$, using the values $\beta_c=12.23$,
  $\nu=0.77$ and $\eta=0.07$. The data approach a scaling curve with
  increasing $L$, supporting the scaling behavior (\ref{chisca}).  }
\label{chin4sca}
\end{figure}

In order to estimate the exponent $\eta$, we analyze the FSS behavior
of the susceptibility given by
\begin{equation}
\chi \sim L^{2-\eta} \left[ f_\chi(X) + O(L^{-\omega})\right].
\label{chisca}
\end{equation}
Fitting the susceptibility to Eq.~(\ref{chisca}) [we use a linear
  approximation for the scaling function $f_\chi(X)$], we obtain the
estimate $\eta=0.07(4)$, where the error takes also into account the
uncertainty on $\beta_c$ and $\nu$. The corresponding scaling plot is
reported in Fig.~\ref{chin4sca}.

The numerical study of the ACP$^3$ lattice model provides therefore a
robust evidence that it undergoes a transition at a finite value of
$\beta$.  We can exclude that the transition is of first order.
Indeed, at a first order transition FSS holds with
$\nu=1/d=1/3$~\cite{NN-75,FB-82,PF-83}.  The estimate (\ref{fitres})
of $\nu$ is definitely larger than $1/3$, ruling out a discontinuous
transition.  Therefore, the transition is
continuous, contradicting the predictions of the LGW
theory~\cite{DPV-15}.

\section{Conclusions}
\label{conclu}

In this paper we have critically reconsidered the LGW approach, which
is used to determine the universal features of critical transitions.
In this framework, one first identifies the order parameter $\Phi$,
then considers the most general $\Phi^4$ theory with the same
symmetries as the original model, and finally determines the stable
fixed points of the RG flow.  If they correspond to a bare theory with
the correct symmetry-breaking pattern, they completely characterize
the possibly present continuous transitions.

In the presence of gauge symmetries---the case of interest here---the
method is usually applied by considering a gauge-invariant order
parameter and a LGW field theory that is invariant under the {\em
  global} symmetries of the original model. In the effective field
theory there is no remnant of the gauge invariance: the gauge degrees
of freedom have been implicitly integrated out. In this work we point
out that in some cases this LGW approach may lead to erroneous
conclusions on the nature of the critical behavior.

As an explicit example we consider the ferromagnetic and
antiferromagnetic CP$^{N-1}$ model, which presents a global U($N$) and
a local U(1) gauge symmetry.  The corresponding LGW theory is
constructed using a hermitian traceless $N\times N$ matrix field
associated with the local gauge-invariant operator $Q_{{\bm x}}^{ab} =
\bar{z}_{\bm x}^a z_{\bm x}^b - \delta^{ab}/N$.

In the ferromagnetic case, for any $N\ge 3$ the LGW theory (\ref{hlg})
contains a cubic term.  Its presence is generally considered as an
indication of first-order transitions.  Therefore, one predicts that
any generic CP$^{N-1}$ model should only undergo discontinuous
transitions. However, this is in contradiction with analytical
large-$N$ results.  In this limit the ferromagnetic CP$^{N-1}$ model
is equivalent to an effective abelian Higgs model for an $N$-component
complex scalar field coupled with a dynamical U(1) field~\cite{MZ-03}.
For a sufficiently large number of components, the abelian Higgs model
has a stable fixed point~\cite{HLM-74}.  Therefore, for $N\to\infty$
CP$^{N-1}$ models may undergo a continuous transition.  The LGW
predictions are also contradicted by recent numerical studies of the
universality class of the 3D ferromagnetic CP$^2$
model~\cite{NCSOS-11,NCSOS-13}.

In the antiferromagnetic case, the LGW field theory is constructed
using the staggered gauge-invariant composite operator, defined as in
Eq.~(\ref{biab}).  It does not present cubic terms due to the
antiferromagnetic nearest-neighbor coupling which gives rise to an
additional global ${\mathbb Z}_2$ symmetry.  For $N=3$, the LGW
approach nicely works: it predicts a symmetry enlargement at the
critical point---the leading critical behavior is O(8)
invariant---which has been accurately verified
numerically~\cite{DPV-15}. In this work we consider the model for
$N=4$.  In this case the LGW predictions are in striking contradiction
with the numerical results.  The analyses of the RG flow presented in
Ref.~\cite{DPV-15} using high-order perturbative series (five-loop
series in the ${\overline {\rm MS}}$ renormalization
scheme~\cite{TV-72} and six-loop series in the massive zero-momentum
scheme~\cite{Parisi-80,ZJ-book,PV-02}) do not find any evidence of
stable fixed points.  This implies that any transition should be of
first order.  On the other hand, the numerical results we present here
provide a robust evidence of a continuous transition in the ACP$^3$
model.

The failure of the LGW approach indicates that the effective local
$\Phi^4$ theory of the gauge-invariant order parameter may not always
capture all the relevant modes at the critical point.  The gauge
degrees of freedom may be relevant at the transition and should
therefore be included in the effective theory.  This is clearly the
case for the large-$N$ ferromagnetic model, which is described by the
abelian Higgs model, in which the U(1) gauge field plays a crucial
role.

A second possibility is that some degrees of freedom decouple giving
rise to continuous transitions associated with different
symmetry-breaking patterns.  For example this occurs in the 2D
frustrated XY models~\cite{HPV-05}, where the disordered
high-temperature phase and the ordered low-temperature phase are
separated by two transitions instead of one, with different critical
modes and symmetry-breaking patterns at each transition (belonging to
the Ising and XY universality classes, respectively).

The above considerations may be relevant for the finite-temperature
transition of quantum chromodynamics (QCD). In the limit of $N_f$
massless quarks, the finite-temperature transition of QCD is related
to the restoring of the chiral symmetry.  The nature of the phase
transition has been investigated within the LGW
framework~\cite{PW-84}, assuming that the relevant order-parameter
field is an $N_f\times N_f$ complex-matrix field $\Phi_{ij}$, related
to the bilinear gauge-invariant quark operators
$\bar{\psi}_{Li}\psi_{Rj}$.  To define the corresponding LGW theory,
one must also specify the fate of the U(1)$_A$ symmetry at the
transition, something which is not clear yet.  Numerical studies of
lattice QCD suggest a strong suppression of U(1)$_A$ symmetry-breaking
effects at $T_c$
\cite{Bazetal-12,CAFHKMN-13,Buchetal-13,AFT-12,BDPV-13}, as predicted
by the dilute instanton gas approximation~\cite{GPY-81}.  In the LGW
approach the role of the axial U(1) symmetry defines the symmetry of
the LGW theory and the relevant symmetry breaking pattern. If the
symmetry is broken, one should consider a $\Phi^4$ model invariant
under ${\rm SU}(N_f)_L\otimes {\rm SU}(N_f)_R$ transformations and the
relevant symmetry breaking pattern is ${\rm SU}(N_f)_L\otimes {\rm
  SU}(N_f)_R \rightarrow {\rm SU}(N_f)_V$; in the opposite case the
symmetry is ${\rm U}(N_f)_L\otimes {\rm U}(N_f)_R$ and the symmetry
breaking pattern is ${\rm U}(N_f)_L\otimes {\rm U}(N_f)_R \rightarrow
{\rm U}(N_f)_V$.  For the particular case of two light flavors, the
two different LGW theories predict two different critical behaviors,
belonging to the O(4) and U(2)$\otimes$U(2) universality classes,
respectively \cite{PV-13}. In any case, whatever the role of the axial
U(1) symmetry is, in the effective LGW theory the gauge symmetry does
not play any role: one is essentially assuming that the gauge degrees
of freedom are irrelevant at the transition.  The results presented in
this paper show that this assumption should not be taken for granted.
It is therefore possible that the problematic consistency and
interpretation of the numerical
results~\cite{CP-PACS-01,MILC-00,KLP-01,KS-01,EHMS-01,DDP-05,KS-06,
  FP-07,Bazetal-12,Burger-etal-13,BDDPS-14,DD-17} is due to the
failure of the LGW framework in describing all critical modes at the
transition. This point calls for further deeper investigations.


\begin{thebibliography}{99}
\bibitem{footnoteMC}
If we set
${\bm Z} = ({\rm Re}\, {\bm z}, {\rm Im}\, {\bm z})$,
the update consists in proposing a new vector
$R\,{\bm Z}$, where $R$ is an O(2) rotation matrix acting on two randomly
chosen components of ${\bm Z}$. The rotation angle is chosen randomly 
in an interval $[-\theta_{\rm max},\theta_{\rm max}]$, where 
$\theta_{\rm max}$ is fixed to have an
acceptance probability of approximately 30\%.
\end{thebibliography}
\end{document}